\documentclass[letterpaper]{article} \usepackage{aaai23}  \usepackage{times}  \usepackage{helvet}  \usepackage{courier}  \usepackage[hyphens]{url}  \usepackage{graphicx}    \usepackage{natbib}  \usepackage{caption}     \usepackage{algorithm}
\usepackage{algorithmic}
\usepackage[implicit=false]{hyperref}

\usepackage{newfloat}
\usepackage{listings}
\DeclareCaptionStyle{ruled}{labelfont=normalfont,labelsep=colon,strut=off} 
\floatstyle{ruled}
\newfloat{listing}{tb}{lst}{}
\floatname{listing}{Listing}
\title{Diversifying Message Aggregation in Multi-Agent Communication via Normalized Tensor Nuclear Norm Regularization}
\author{
	Yuanzhao Zhai\textsuperscript{\rm 1}, Kele Xu\textsuperscript{\rm 1}, Bo Ding\textsuperscript{\rm 1}\thanks{Corresponding authors: Kele Xu and Bo Ding}, Dawei Feng\textsuperscript{\rm 1}, Zijian Gao\textsuperscript{\rm 1}, Huaimin Wang\textsuperscript{\rm 1} \\
}
\affiliations{
	\textsuperscript{\rm 1} National University of Defense Technology, Changsha 410005, China\\
	yuanzhaozhai@nudt.edu.cn,
	kelele.xu@gmail.com, 
	dingbo@nudt.edu.cn, \\
	davyfeng.c@gmail.com, 
	gaozijian19@nudt.edu.cn, 
	whm\_w@163.com \\				
}

\usepackage{bibentry}

\usepackage{bibentry}

\usepackage[tight,footnotesize]{subfigure}
\usepackage[mathscr]{eucal}

\usepackage{amsmath}
\usepackage{amssymb}
\usepackage{bm}

\usepackage{indentfirst}
\usepackage{amsmath}

\usepackage{makecell,rotating,multirow,diagbox}

\usepackage{cleveref}

\begin{document}
	
	\maketitle
	
	\begin{abstract}
		Aggregating messages is a key component for the communication of multi-agent reinforcement learning (Comm-MARL). Recently, it has witnessed the prevalence of graph attention networks (GAT) in Comm-MARL, where agents can be represented as nodes and messages can be aggregated via the weighted passing. While successful, GAT can lead to homogeneity in the strategies of message aggregation, and the ``core'' agent may excessively influence other agents' behaviors, which can severely limit the multi-agent coordination. To address this challenge, we first study the adjacency tensor of the communication graph and demonstrate that the homogeneity of message aggregation could be measured by the normalized tensor rank. Since the rank optimization problem is known to be NP-hard, we define a new nuclear norm, which is a convex surrogate of normalized tensor rank,  to replace the rank. Leveraging the norm, we further propose a plug-and-play regularizer on the adjacency tensor, named \emph{Normalized Tensor Nuclear Norm Regularization} (NTNNR), to actively enrich the diversity of message aggregation during the training stage. We extensively evaluate GAT with the proposed regularizer in both cooperative and mixed cooperative-competitive scenarios. The results demonstrate that aggregating messages using NTNNR-enhanced GAT can improve the efficiency of the training and achieve higher asymptotic performance than existing message aggregation methods. When NTNNR is applied to existing graph-attention Comm-MARL methods, we also observe significant performance improvements on the StarCraft II micromanagement benchmarks.

													\end{abstract}

	\section{Introduction}
	\noindent 
	Multi-Agent Reinforcement Learning (MARL) has achieved remarkable success in a range of challenging sequential decision-making tasks, such as traffic control~\cite{zhou2020smarts}, swarm robotics~\cite{zhai2021decentralized} and multi-player strategy games~\cite{yuan2022multi}.
		As an under-explored issue in MARL, communication is a key component for multi-agent coordination where agents can exchange their local observations via communication messages. These messages are aggregated by decentralized agents and further utilized to augment individual local observations for learning policies and selecting actions, allowing the agents to jointly optimize the objectives. 
	
	\begin{figure*}[t]
		\centering
		\subfigure[With homogeneous message aggregation strategies obtained by GAT, predators 2, 3, and 4 are excessively influenced by the message of predator 1.
		All predators tend to pursue prey 1 while ignoring prey 2.
				\label{fig-ppa}]
		{
			\centering
			\includegraphics[width=0.47\linewidth]{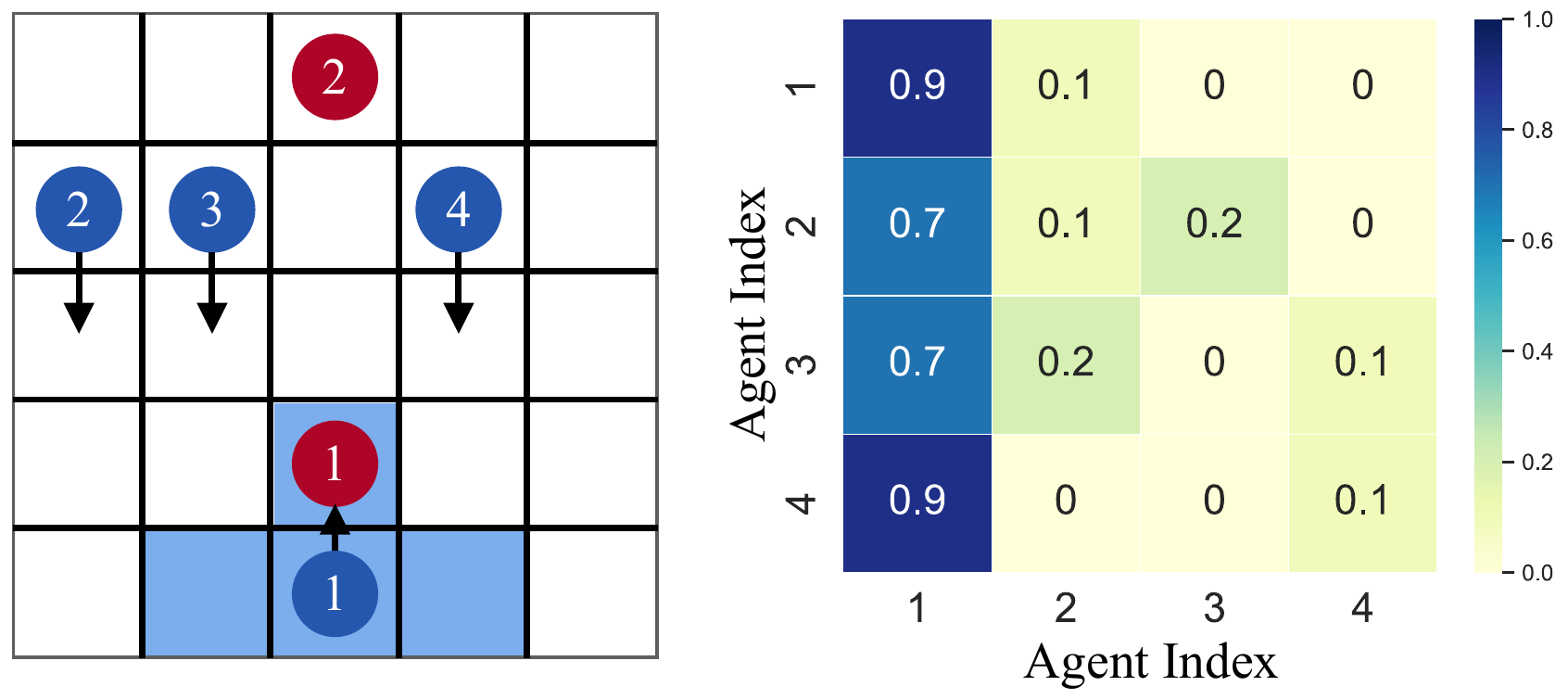}
		}
		\quad
		\subfigure[With diverse message aggregation strategies, predators 1 and 2 intend to capture prey 1, while predators 3 and 4 explore the environment. In this way, the predators could obtain better coordination.
		\label{fig-ppb}]
		{
			
			\centering
			\includegraphics[width=0.47\linewidth]{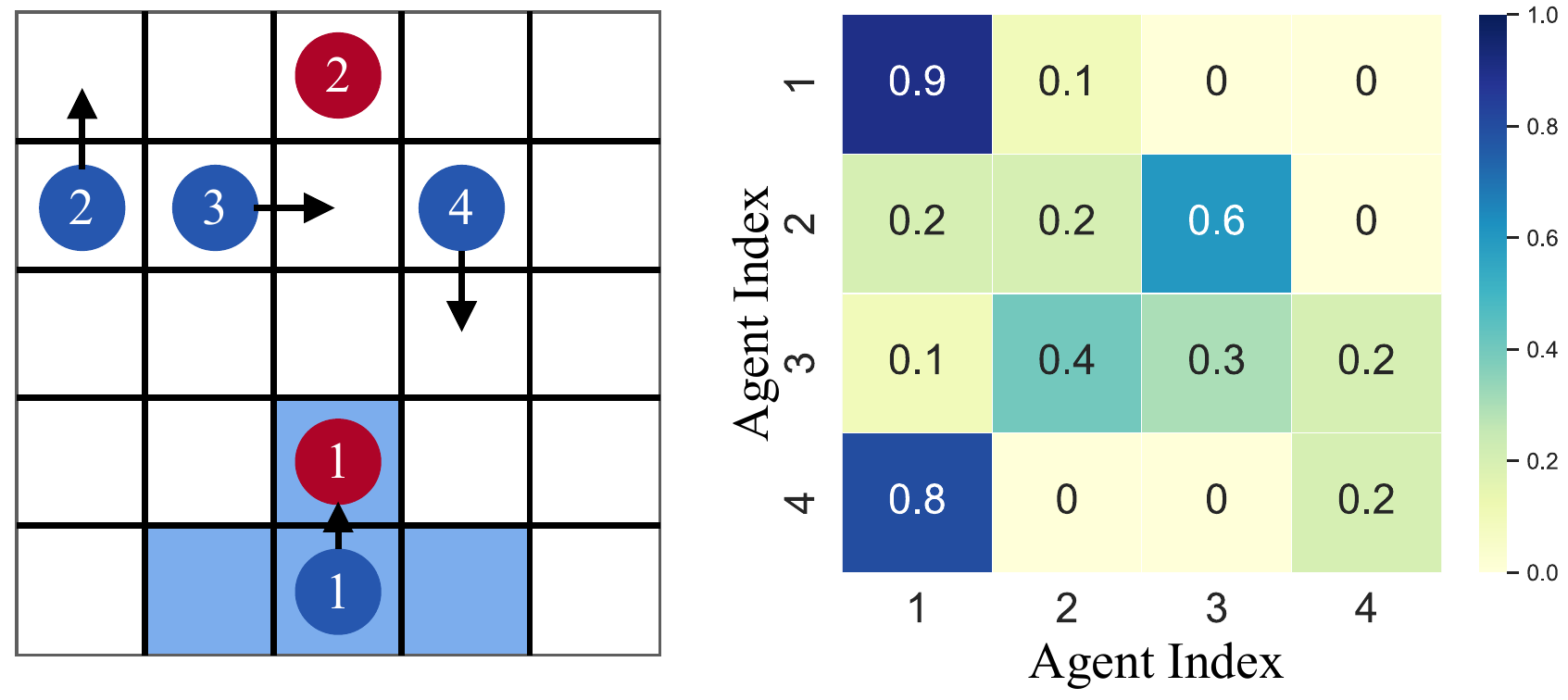}
		}
		\caption{A toy experiment in the predator-prey scenario. 
			Predators are marked in blue, and preys are in red.
			Four predators with a limited vision of one grid are pursuing two static preys through communication.
			Exact two predators are required to be present in the grid cell of a prey for a successful capture.
			Element at row $i$ and column $j$ of the adjacency matrix represent the attention score of agent $j$'s communication message to agent $i$.
						\label{fig-pp}}
		
	\end{figure*}

	Although sustainable efforts have been made, efficient communication between agents is still far from being solved. 
		How to aggregate messages is a key factor that determines communication efficiency.
			To model the interactions between agents, MARL has widely utilized graph neural networks (GNNs)~\cite{scarselli2008graph} to allow for a graph-based representation.
				The multi-agent system is usually modeled as a complete graph, and each agent corresponds to a node.
			As one of the most popular GNNs variants, GAT has shown great potential in Comm-MARL~\cite{zhu2022survey}.
	Message aggregation can be achieved via attention-weighted message passing in the communication graph.

	Despite the success of the GAT in Comm-MARL, we show that a lack of diversity still persists in the obtained message aggregation strategy.
		In essence, many nodes in the graph may pay undue attention to a few ``key'' nodes and are often excessively influenced.
	This issue is identified in various tasks modeled by GAT~\cite{brody2022attentive}, and multiple agents exacerbate the problem severely.
		For the multi-agent scenarios where the importance of messages is conditioned on agents' state, homogeneous message aggregation strategies mean most agents may pay excessive attention to some emergent message, resulting in inefficient communication.
	Moreover, since many Comm-MARL methods adopt the parameter-sharing scheme, agents with homogeneous message aggregation strategies tend to obtain similar behaviors, severely limiting the diversity of behaviors for better coordination~\cite{chenghao2021celebrating}.
		As shown in Figure~\ref{fig-pp}, the behavior obtained by methods with homogeneous message aggregation strategies can be suboptimal, highlighting the urgent need for diverse message aggregation strategies.
	
	In this paper, we aim to enable agents to explore diverse message aggregation strategies.
	Firstly, we study the adjacency tensor of the multi-agent communication graph, which consists of adjacency matrices generated by the multi-head attention mechanism of GAT.
	We present that the homogeneity of message aggregation could be measured by the normalized tensor rank and normalized tensor nuclear norm.
	Accordingly, we propose a novel \emph{Normalized Tensor Nuclear Norm} (NTNN) regularizer, which regularize adjacency tensors to actively enrich the diversity of the message aggregation strategies in Comm-MARL. 
		In this way, agents could discover diverse behaviors and tend to find better coordination. In brief, our main contribution is threefold:
	\begin{itemize}
		\item We firstly propose to measure the diversity (or the homogeneity) of the message aggregation via the normalized tensor rank of the adjacency tensor.
				\item We define a novel normalized tensor nuclear norm to replace the rank. The norm can be further utilized as the regularizer to discover diverse message aggregation strategies for multi-agent communication.
						\item Experiments show that aggregating messages using GAT with NTNNR can improve training efficiency and asymptotic performance.
		Our regularizer also brings significant performance improvements for existing graph-attention Comm-MARL methods, using the plug-and-play manner.
	\end{itemize}

	\section{Related Work}{\label{related}}
	\subsection{Attention in Graphs}
			\noindent 
	For graph-structured data, attention mechanisms have been widely used to model the pairwise interactions between nodes.
	For example, many previous attempts employed GNNs with attention mechanisms~\cite{lee2019attention,brody2022attentive}, which generalizes the standard node representation update pattern, e.g., averaging or max-pooling of neighbors~\cite{kipf2017semi, hamilton2017inductive}.
	During the message passing, the attention mechanism allow nodes to compute a weighted average of their neighbors, and softly select their most relevant neighbors.
	GAT~\cite{velivckovic2018graph} is one of the most popular GNNs variants. GAT generalizes the multi-head self-attention mechanism~\cite{vaswani2017attention} from sequences to graphs, which allows the model to attend to information from different representation subspaces jointly.
	
	Despite the effectiveness, GATv2~\cite{brody2022attentive} finds by a theoretical analysis that the ranking of the attention scores generated by GAT may be unconditioned on the query node, which is called the static attention problem.
			To address this problem, GATv2 proposes to modify the order of weight calculation operation in GAT, outperforming GAT in many public datasets of GNNs.
	
	\subsection{Message Aggregation Methods in Comm-MARL}
	\noindent 
	In Comm-MARL, message aggregation strategies for agents determine how to aggregate received messages and partial observation to select the next actions.
	Some works aggregate messages with no preference, such as concatenation~\cite{foerster2016learning, kim2019learning, kim2020communication}, averaging~\cite{sukhbaatar2016learning, singh2019individualized}, summing up~\cite{du2021learning}, recurrent neural networks~\cite{peng2017multiagent}, and so on.
	Since messages encode the senders' personal understanding of their observations, some may be more important than others.
	
	To aggregate messages unequally, the attention mechanism is often utilized to calculate received messages' weights and then aggregate them together~\cite{das2019tarmac, agarwal2020learning}.
	Considering the graph topology, GAT has been proved an effective tool to aggregate messages~\cite{liu2020multi,li2021deep, niu2021multi}.
	GA-Comm~\cite{liu2020multi} propose a two-stage graph-attention mechanism.
	The hard attention determines whether communication between agents is necessary, while the soft attention calculates the attention weight.
		DICG~\cite{li2021deep} introduces the deep implicit coordination graph with the self-attention mechanism for message aggregation.

	Inherited from the static attention problem, most methods mentioned above lack diversity in terms of message aggregation.
	Although replacing GAT with GATv2 in Comm-MARL can prove the diversity of message aggregation theoretically, agents with similar observation still tend to obtain homogeneous message aggregation strategies in practice.
		Complementary with GATv2, our method regularize the adjacency tensor of GAT, encouraging diverse message aggregation actively.

				\subsection{Diversity in MARL}
	\noindent 
	As an emerging topic in MARL, maintaining diversity for policies is meaningful for various scenarios, such as emergent behavior~\cite{tang2020discovering}, exploration~\cite{mahajan2019maven} or learning to adapt~\cite{balduzzi2019open}.
		Diverse policies can be discovered by evolution methods~\cite{cully2015robots,pugh2016quality}, specially designed reward function~\cite{lowe2019pitfalls,baker2019emergent,tang2020discovering}, role-based learning~\cite{wang2019influence,wang2021rode}, population-based training~\cite{vinyals2019grandmaster,parker2020effective,lupu2021trajectory} or iterative policy optimization~\cite{zhou2021continuously,zahavy2021discovering}.

	Based on the value decomposition framework, some attempts aim to maintain diversity through non-shared individual Q-functions for each agent.
	EOI~\cite{jiang2021emergence} combines the gradient from the intrinsic value function (IVF) and the total Q-function to train each agent's local Q-function.
	CDS~\cite{chenghao2021celebrating} maximize the mutual information between agents' identities and their trajectories to diversify individual Q-functions.
	However, a recent work~\cite{fu2022revisiting} theoretically shows that policy gradient with individual policy or communication can be comparable to popular value-based learning methods for maintaining diverse policies.
		They propose to obtain diverse policies with an auto-regressive policy gradient. Each agent selects actions according to different other agents' actions, which can be seen as a Comm-MARL method.
		
	Most of the aforementioned methods did not study diverse message aggregation strategies, which can be a potential way to enrich diversity in Comm-MARL.
		We adopt the policy gradient method with parameter sharing and utilize GAT with NTNNR to aggregate messages.
	To our best knowledge, we are the first to discover diverse policies through diversifying message aggregation.

	\section{Background and Notations}
	\noindent 
	We model the multi-agent tasks as a Decentralized 
	Partially Observable Markov Decision Process (Dec-POMDP) augmented with communication, which can be described as a tuple $<N, \mathcal{S}, \mathcal{U}, \mathbf{P}, \mathbf{R}, \mathcal{O},\mathcal{M},\mathcal{G},\gamma>$. 
			$N$ is the number of agents. 
	$\mathcal{S}$ represents the space of global states. 
	$\mathcal{O}$ denotes the space of observations of robots, and each agent receives a private observation $o_i \in \mathcal{O}$ according to the observation function $\sigma(\mathbf{s}_i): \mathcal{S} \rightarrow  \mathcal{O}$.
	$\mathcal{M}$ represents the space of messages.
	Agents generate messages ${m}_i \in \mathcal{M}$ encoded by its observations and others' messages at the last timestep, which could be modeled by the multi-agent communication graph $\mathcal{G} = <\mathcal{V},\mathcal{E}>$.
	Node $v_i \in \mathcal{V}$ represent agents, and edges $e_{ij }\in \mathcal{E}$ represent communication links.
	We denote $h_i$ as the feature of $v_i$.
				Combining with local observations, each agent aggregates communication messages and generates its own action $u_i = \pi_{\boldsymbol{\theta}}(o_i, m_{j \neq i})$, where $\pi_{\boldsymbol{\theta}} $ is the policy with parameter $\boldsymbol{\theta}$ shared across all agents.
			For states $s,s^\prime \in \mathcal{S}$ and a joint action $\mathbf{u} \in \mathcal{U}^N$, the transition probability of reaching state $s^\prime$ from state $s$ by executing action a is $\mathbf{P}(s^\prime |s,\mathbf{u})$.
			$\mathbf{R}$ is the joint reward function.
			$\gamma \in [ 0,1 ] $ denotes the discount factor.
	Agent $i$ aims to maximize its discounted reward 
	$
		\mathbb{E}_{s \sim \rho_{\pi},\mathbf{u} \sim {\pi}} [{r}_i^t] =  \sum_{t=0}^{\infty} \gamma ^t r_i^t(s^t, \mathbf{u}^t) $, where $\rho_{ \pi}$ is the discounted state distribution induced by the policy.
	
		In this paper, we denote adjacency tensors by boldface Euler script letters $\mathcal {A}$. Adjacency matrices are denoted by boldface capital letters $\mathbf {A}$; vectors are denoted by boldface lowercase letters $\boldsymbol a$, and scalars are denoted by lowercase letters $a$.
	For the communication graph of GAT, adjacency matrices $\mathbf {A} \in \mathbb{R}_+^{N \times N}$ generated by the multi-head attention mechanism can be regarded as a three-way adjacency tensor $\mathcal {A} \in \mathbb{R}_+^{N \times N \times K}$, where the dimension of the third way is the number of attention heads $K$.
	We denote the $(i, j, k)$-th entry of $\mathcal {A}$ as $\mathcal { A}_{ijk} $.
	The frontal slice $\mathcal A(:, :, k)$ is denoted compactly as $\mathbf A ^{(k)}$.

	\section{Methodology}
	\label{Sec:methods}
\begin{figure}[t]
		\centering
		\includegraphics[width=1\linewidth]{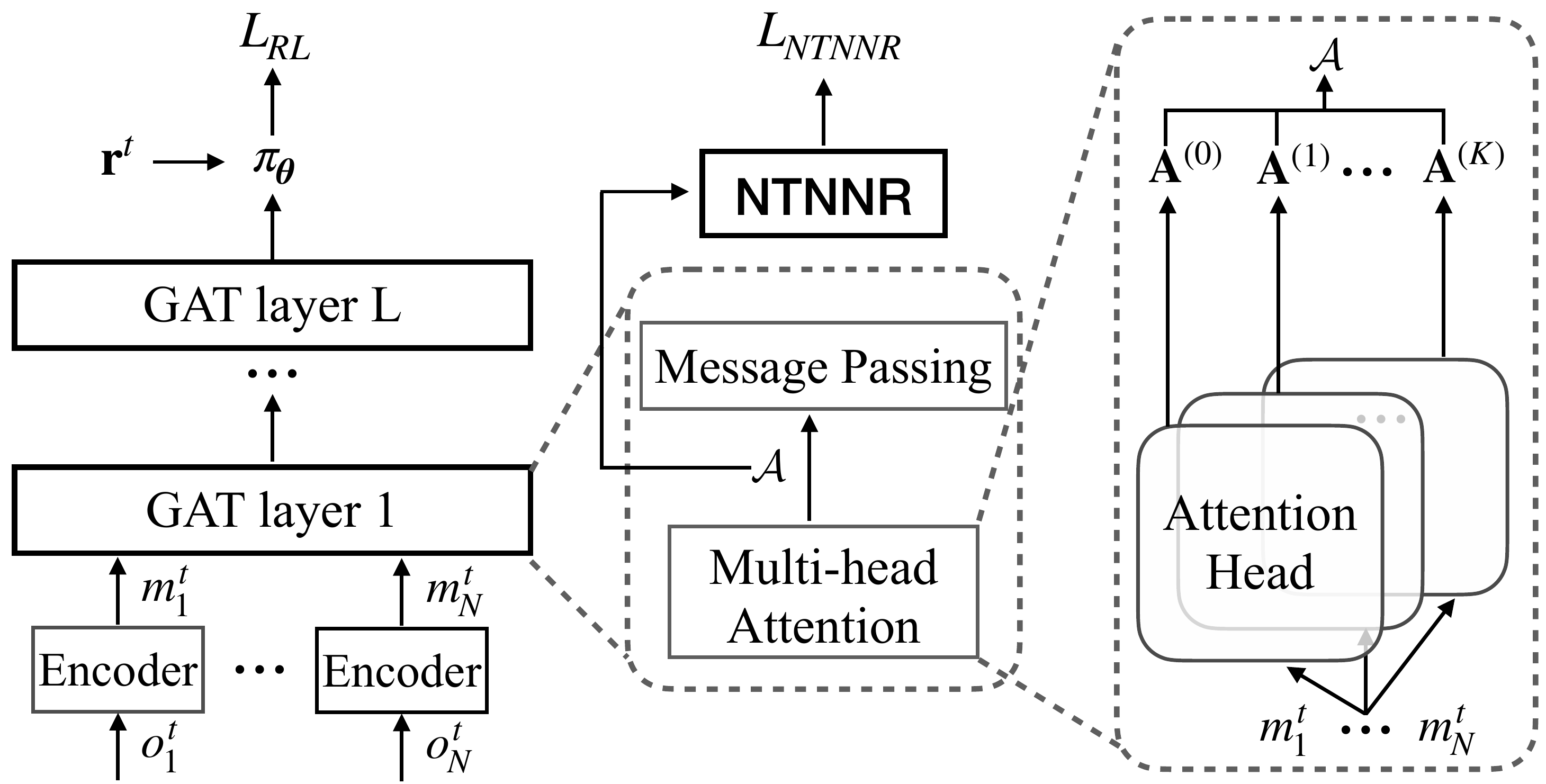}
		\caption{
						Schematics of NTNNR.
			We regularize the NTNN of the adjacency tensor $\mathcal{A}$ which consists of adjacency matrices generated by the multi-head attention mechanism.
			\label{fig:NTNNR}}
	\end{figure}

	\noindent 
	In this section, we describe the details of NTNNR (Figure~\ref{fig:NTNNR}), which actively enrich the diversity of message combination and can be integrated into graph-attention Comm-MARL methods.
				
	\subsection{Measuring Message Aggregation's Diversity with the Normalized Tensor Rank}
	\noindent 
	For each agent $i$, GAT computes a learnable weighted average of the representations of all neighbors $j \in \mathcal{N}_i$. 
	\begin{equation}
		e(\mathbf{\mathbf h_i},\mathbf{\mathbf h_j}) =  LeakyReLU (\mathbf {W'} [\mathbf {W} \mathbf h_i \Vert \mathbf {W} \mathbf h_j]),
	\end{equation}
	where $\mathbf {W}$ and $\mathbf {W'}$ are learnable, and $\Vert$ denotes vector concatenation.
	
	We first consider the case that a single attention head is used. Then the attention scores, as the elements of the adjacency matrix $\mathbf A$, are normalized across all neighbors using the softmax function:
	\begin{equation}
		a_{ij} = Softmax_j (e(\mathbf{\mathbf h_i},\mathbf{\mathbf h_j})) = \frac{exp({e(\mathbf{\mathbf h_i},\mathbf{\mathbf h_j})})}{\sum_{j' \in \mathcal{N}_i } exp(e(\mathbf{\mathbf h_i},\mathbf{\mathbf h_j'}))}.
	\end{equation}
	
	The adjacency matrix $\mathbf A \in \mathbb{R}_+^{N \times N}$ satisfies the following properties:
							
	\begin{equation}
		\left\{
		\begin{aligned}
			\ \ \ \ \sum_{j=1}^{N} &a_{ij}  = 1 \ \ \ \   \forall i \in 1,\cdots, N,
			\\& a_{ij} \geq 0 \ \ \ \   \forall i \in 1,\cdots, N, j \in 1,\cdots, N.
		\end{aligned}
		\right.
	\end{equation}
	
	We denote the vectors selected from the $i$-th and $j$-th rows of the matrix as $\boldsymbol a_i$ and $\boldsymbol a_j$, which represent the attention scores of agent $i$ and $j$ respectively for aggregating messages.
	If agent $i$ and $j$ have homogeneous message aggregation strategies, the difference between $\boldsymbol a_i$ and $\boldsymbol a_j$ is minor.
	In this case, $\boldsymbol a_i$ and $\boldsymbol a_j$ could be approximately regarded as linearly dependent.
	On the contrary, diverse message aggregation strategies mean linearly independent vectors.
	Therefore, we could measure the diversity (or the homogeneity) of the message aggregation with the matrix rank of the adjacency matrix $\mathbf {A}$.
	
	\begin{figure}[t]
		\subfigure[Frontal slices view.
		\label{fig-slice}]
		{
			\centering
			\includegraphics[width=0.44\linewidth]{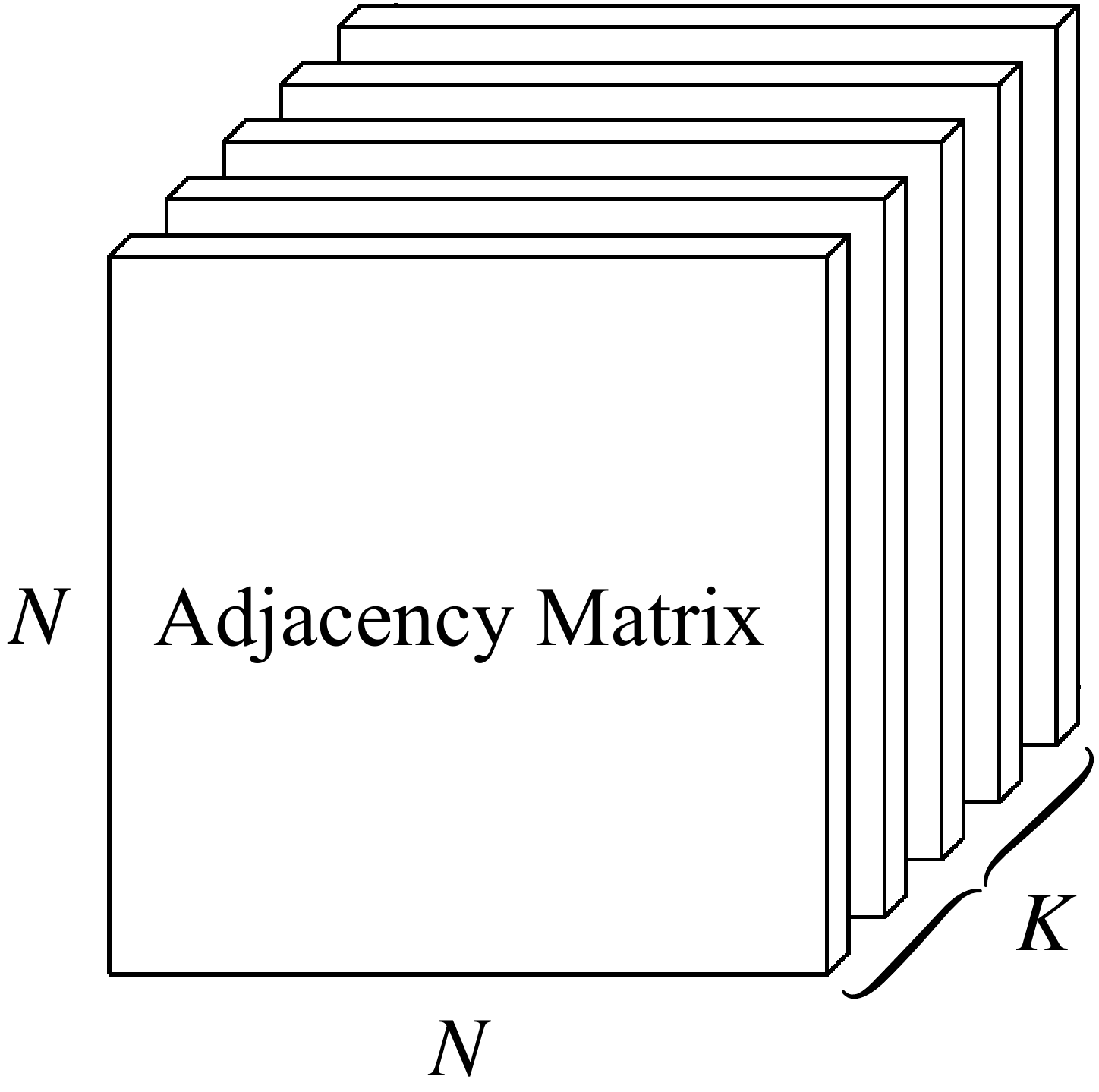}
		}
		\subfigure[Mode-3 fibers view.
		\label{fig-fiber}]
		{
			\centering
			\includegraphics[width=0.46\linewidth]{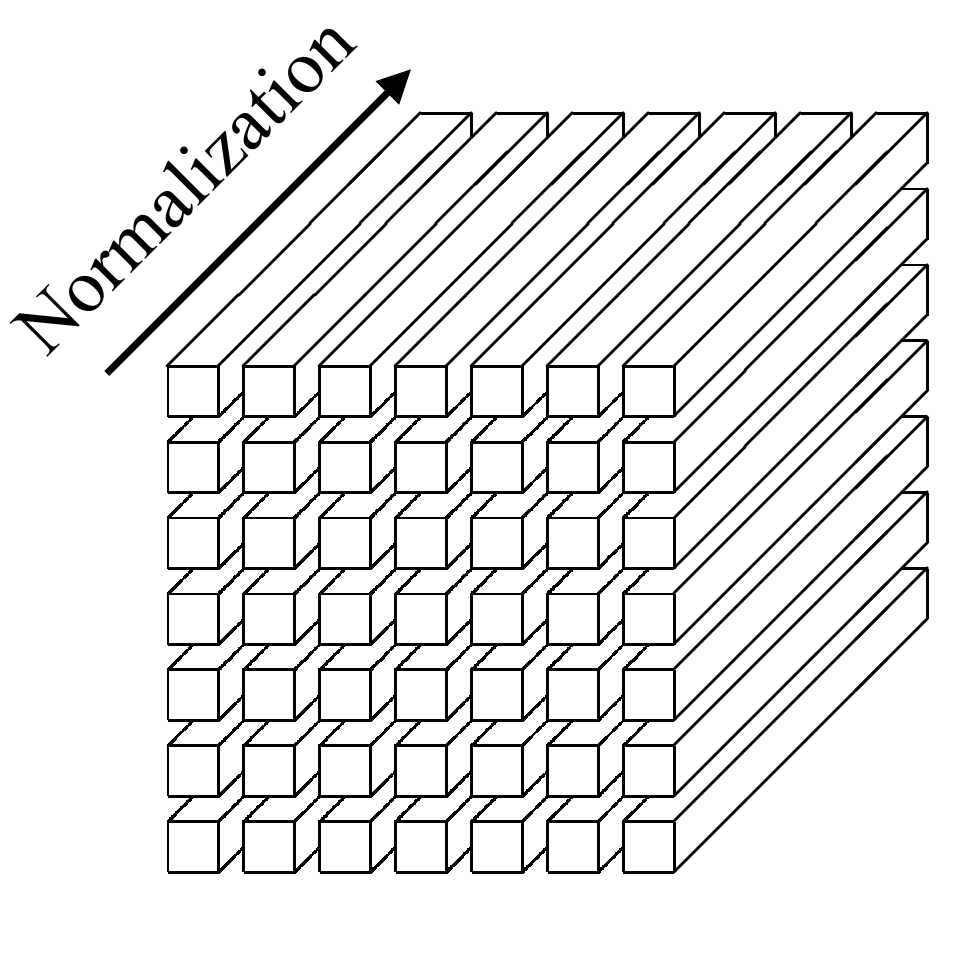}
		}
		
		\caption{Multiple views of the adjacency tensor $\mathcal A$. \label{fig-tensor}}
		
	\end{figure}
	
	With the multi-head attention mechanism in GAT, independent $K$ attention mechanisms execute the attention function in parallel. Then we obtain a three-way adjacency tensor $\mathcal A \in \mathbb{R}_+^{N \times N \times K}$.
	As shown in Figure~\ref{fig-slice}, $K$ frontal slices $ \{\mathbf A ^{(k)} \}_{i=1,\cdots,K}$ represent independent adjacency matrices.
	From another view, as shown in Figure~\ref{fig-fiber}, the mode-3 fiber $\mathcal A(i, j, :)$ represents the attention scores of agent $j$ to agent $i$ using different attention heads.
	Multiple heads are considered to attend to information from different representation subspaces.
	Thus we aim to maintain the diversity in both the frontal slices view and the mode-3 fibers view of the adjacency tensor.
	
	Extended from the matrix rank, tensor rank could be defined in various ways.
	CP rank~\cite{kolda2009tensor} denotes the smallest number of rank one tensor decomposition. But both CP rank and its convex relaxation is hard to obtain.
		To avoid this issue, the tractable Tucker rank~\cite{kolda2009tensor} and its convex relaxation are more widely used.
				However, most existing tensor ranks can not directly measure the linear correlation from both frontal slices and mode-3 fibers views.
	This motivates us to define a new tensor rank to measure the homogeneity of message aggregation with multi-head attention GAT.
	
	We denote $ \mathcal{\hat A}$ as a result of applying normalization to $\mathcal{A}$ along the 3-rd way.
				Specifically, we apply Softmax on every tube fibers $ \mathcal {A}(i, j, :)$, i.e.,
	\begin{equation}
		\mathcal {\hat A}_{ijk}  = \frac{exp(\mathcal {A}_{ijk})}{\sum_{l \in [0, K -1]} exp(\mathcal {A}_{ijl})} , \ \ K \geq 2.
		\label{normlization}
	\end{equation}
	
	Then we can define the normalized tensor rank as:
	\begin{equation}
		\text{rank}_n(\mathcal A) = \sum_{k} \text{rank}( \mathbf{\hat A}^{(k)}).
		\label{rank}
	\end{equation}

	\subsection{Normalized Tensor Nuclear Norm Regularization}
	\noindent 
	The rank optimization problem is known to be NP-hard. An alternative is to utilize the nuclear norm, and the matrix nuclear norm is defined as:
	\begin{equation}
		\Vert \mathbf {A} \Vert _ * = \sum_{i} \sigma_i (\mathbf {A}),
	\end{equation}
	where $\sigma_i (\mathbf {A})$ are singular values of $ \mathbf {A}$. 
	
	For normalized tensor $ \mathcal{\hat A}$, we denote $\mathbf {\hat A} \in \mathbb{R}_+^{NK \times NK}$ as the block diagonal matrix with its $i-th$ block on the diagonal as the $i-th$ frontal slice, i.e., 
	\begin{equation}
		\mathbf {\hat A} = bdiag(\mathcal{\hat A})= \begin{bmatrix}  \mathbf {\hat A }^{(0)} & &   \\ &   \mathbf {\hat A }^{(1)}& \\ && \ddots \\ &  & &\mathbf {\hat A }^{(K -1)}\end{bmatrix}.
		\label{tensor-matrix}
	\end{equation}
	
	Based on the matrix nuclear norm, we define a novel tensor nuclear norm for the normalized tensor rank, which is called \emph{Normalized Tensor Nuclear Norm} (NTNN):
	\begin{equation}
		\Vert \mathcal{A} \Vert _ * =\frac{1}{K}\Vert \mathbf {\hat A} \Vert _ *.
	\end{equation}
	
		As a special case, if $\mathcal{A}$ reduces to a matrix ($K = 1$),  it is not necessary to normalize the third dimension. 
	In this case, NTNN reduces to the matrix nuclear norm.
		Considering the nuclear norm is the convex relaxation of the matrix rank~\cite{candes2009exact}, $\Vert \mathbf{\hat A} \Vert_ *$ is a tight convex surrogate of $\text{rank}(\mathbf{\hat A}) $.
	Combining Equation~\ref{rank} and \ref{tensor-matrix}, $\Vert \mathcal{A} \Vert _ *$ is a tight convex surrogate of $\text{rank}_n(\mathcal A) $.

				Regularizing NTNN of the adjacency tensor $\mathcal{A}$ could maintain the diversity of message aggregation.
			With the increase of NTNN, the diversity of $\mathcal{A}$ is enriched not only in the frontal slices view but also in the mode-3 fibers view, which makes agents' message aggregation strategies more diverse.

	\begin{figure*}[ht]
														\centering
		\subfigure[Average reward of various message aggregation approaches.
		\label{fig-pp-reward}]
		{
			\centering
			\includegraphics[width=0.305\linewidth]{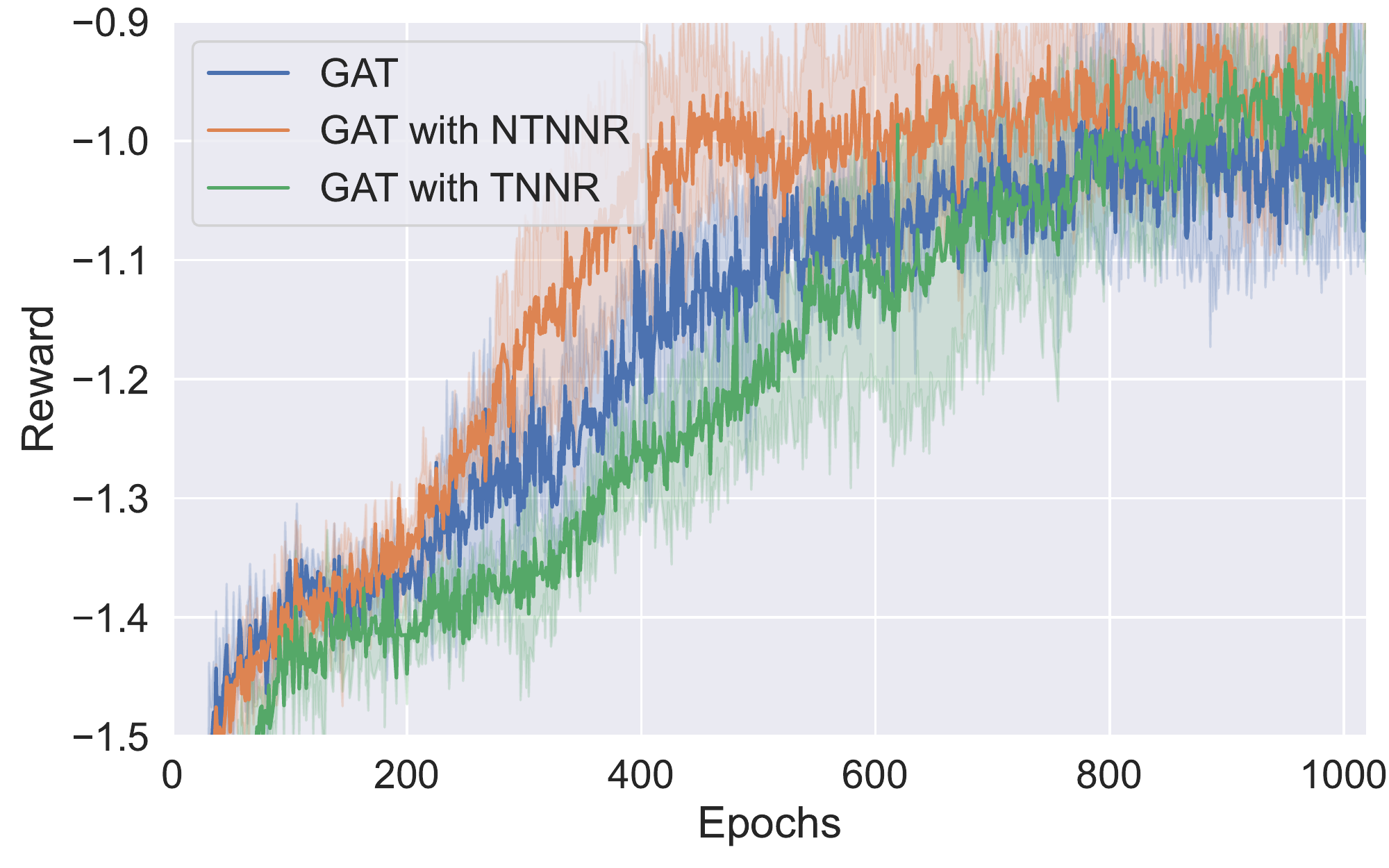}
		}
		\quad
		\subfigure[Value of NTNN of GAT layer $1$.
		\label{fig-pp-NTNNR-layer1}]
		{
			\centering
			\includegraphics[width=0.305\linewidth]{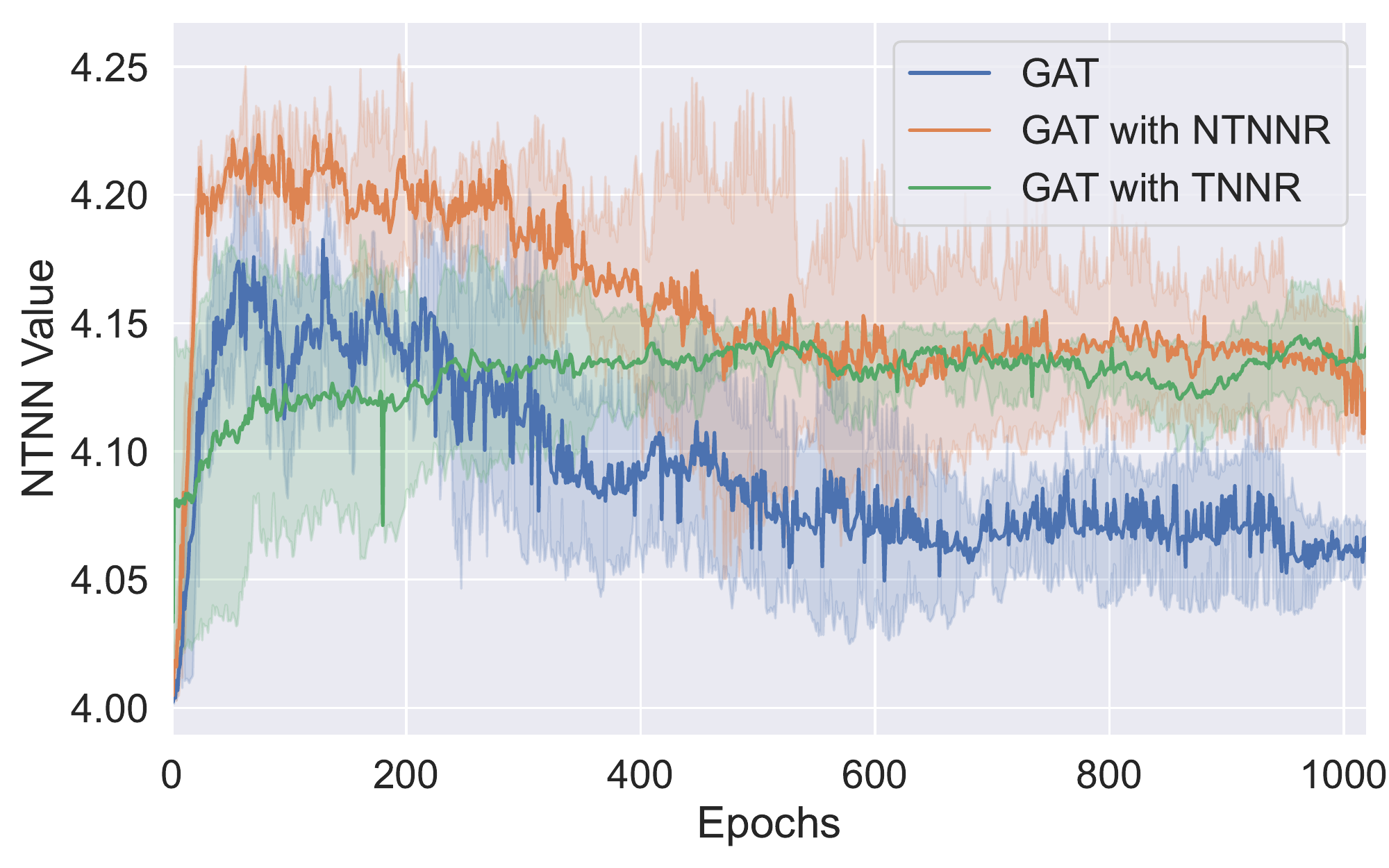}
		}
		\quad
		\subfigure[Value of NTNN of GAT layer $2$.
		\label{fig-pp-NTNNR-layer2}]
		{
			\centering
			\includegraphics[width=0.305\linewidth]{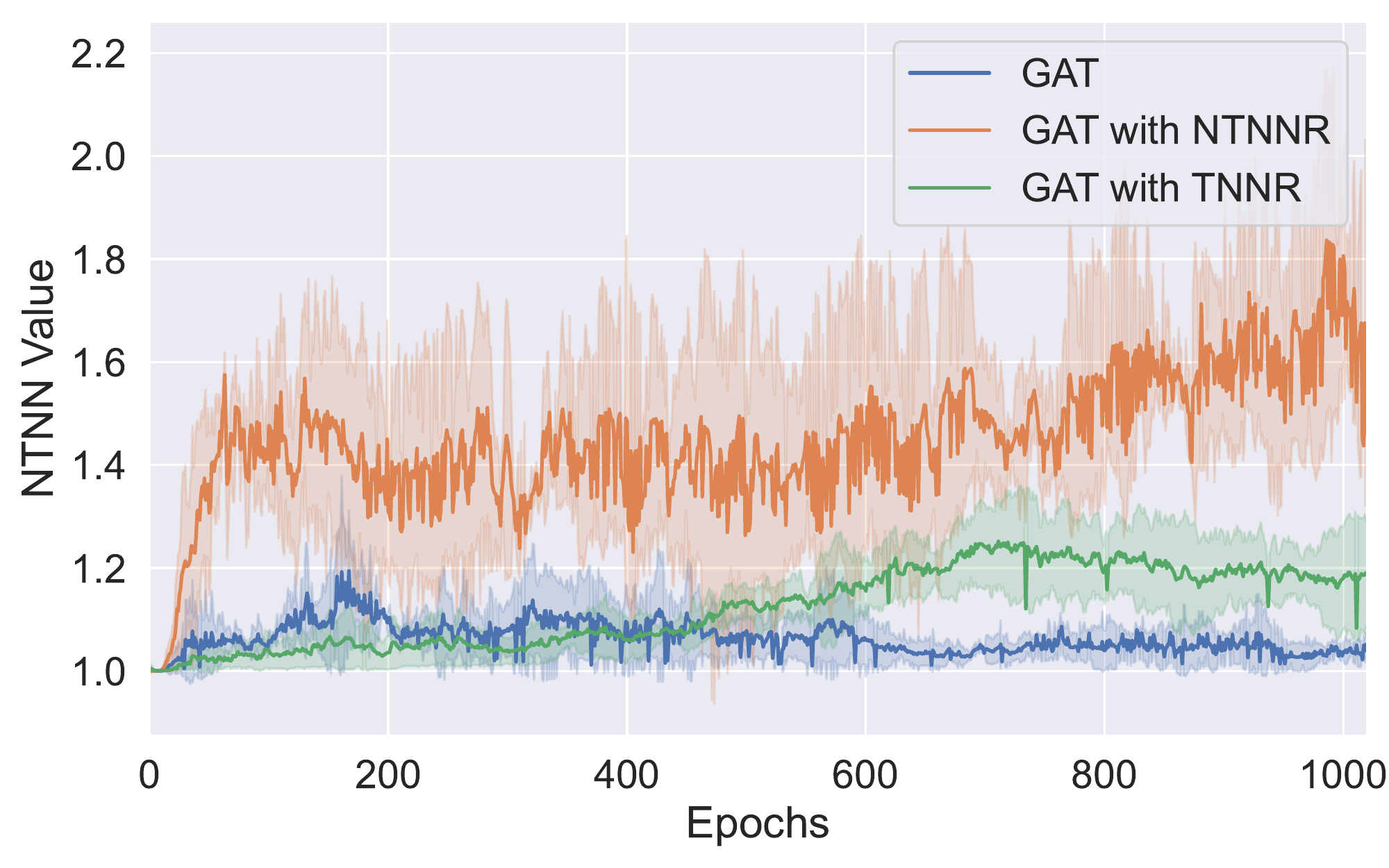}
		}
														\caption{
			A case study in the predator-prey scenario indicates that NTNNR encourages diverse policies, thus achieving higher training efficiency and asymptotic performance.
			\label{fig-pp-NTNNR}}
	\end{figure*}

	\subsection{Overall Optimization Objective}
	\noindent 
	In this part, we describe how to use NTNNR to diversify message aggregation strategies in graph-attention Comm-MARL algorithms.
	
	\begin{algorithm}[tb]
		\caption{Comm-MARL with NTNNR}
		\label{alg:algorithm}
		\textbf{Initialization}: the number of agents $N$ ,  the number of the communication graph layers $L$, parameters of the policy network ${\boldsymbol{\theta}}$
		\begin{algorithmic}[1] 						\WHILE{Training}
			\STATE ${L}({\boldsymbol{\theta}})  \leftarrow 0$
			\FOR {each agent $n \in range(N)$}
			\STATE Calculate the Comm-MARL loss ${L}_{RL}({\boldsymbol{\theta}})$
			\STATE ${L}({\boldsymbol{\theta}})  \leftarrow {L}({\boldsymbol{\theta}})  + {L}_{RL}({\boldsymbol{\theta}})$
			\ENDFOR
			\FOR {each communication graph layer $l\in range(L)$}
			\IF {Number of attention heads is greater than 1}
			\STATE Normalize the adjacency tensor as Equation~\ref{normlization}.
			\ENDIF
			\STATE ${L}_{NTNNR}(\boldsymbol{\theta}_l)  =-\Vert \mathcal{A} \Vert _ *$
			\STATE Calculate the $\lambda_l$ as Equation~\ref{lambda}.
			\STATE ${L}({\boldsymbol{\theta}})  \leftarrow {L}({\boldsymbol{\theta}})  + \lambda_l {L}_{NTNNR}(\boldsymbol{\theta}_l) $
			\ENDFOR
						\STATE ${\boldsymbol{\theta}}  \leftarrow  optimize({L}({\boldsymbol{\theta}}))$
			\ENDWHILE
					\end{algorithmic}
	\end{algorithm}
	
	Following most Comm-MARL methods, we implement our framework with the policy decentralization with shared parameters (PDSP) paradigm.
					Then the gradient of Comm-MARL's original loss function can be formulated as:
								
	\begin{equation}
		\nabla_{\boldsymbol{\theta}}	{L}_{RL}(\boldsymbol{\theta})  =  \mathbb{E}_{i,t} [\nabla_{\boldsymbol{\theta}}log \  \pi_{\boldsymbol{\theta}}(u_i^t | o_i^t, m_{j \neq i}^t) {\Psi}_i^t ],
			\end{equation}
	where ${\Psi}_i^t $ is related to the discounted reward ${r}_i^t $ and has various forms depending on different algorithms~\cite{schulman2015high}, and ${\boldsymbol{\theta}}$ denotes all parameters of the policy network.

		To discover diverse message aggregation strategies, we apply NTNNR to the adjacency tensor $\mathcal{A}$ of GAT layers. The corresponding loss function of NTNNR in the $l$-th layer can be formulated as:
	\begin{equation}
		{L}_{NTNNR}(\boldsymbol{\theta}_l)  =-\Vert \mathcal{A} \Vert _ *,
	\end{equation}
	where $\boldsymbol{\theta}_l$ is part of parameters ${\boldsymbol{\theta}}$ to obtain the adjacency tensor $\mathcal{A}$ of the $l$-th GAT layer.
	
	Overall, we update the model parameter $\boldsymbol{\theta}$ by minimizing the following loss function:
	\begin{equation}
		{L}({\boldsymbol{\theta}}) = {L}_{RL}({\boldsymbol{\theta}})+ \sum_{l} \lambda_l {L}_{NTNNR}(\boldsymbol{\theta}_l),
	\end{equation}
	where $\lambda_l$ is the regularization weights of NTNNR for layer $l$.
	To anneal $\lambda_l$ during the training process, we introduce new scaling hyper-parameters $\beta_l$ and obtain adaptive weight as follows:
	\begin{equation}
		\lambda_l  = \frac{|{L}_{RL}(\boldsymbol{\theta})|}{\beta_l \times | {L}_{NTNNR}(\boldsymbol{\theta}_l)| } 
		\label{lambda}.
	\end{equation}
	
	Algorithm 1 details how NTNNR is integrated with generic Comm-MARL algorithms.

	\begin{figure*}[ht]
														
		\subfigure[The frontal slices with TNNR.
		\label{fig-pp-TNNR-slice0}]
		{
			\includegraphics[width=0.5\linewidth]{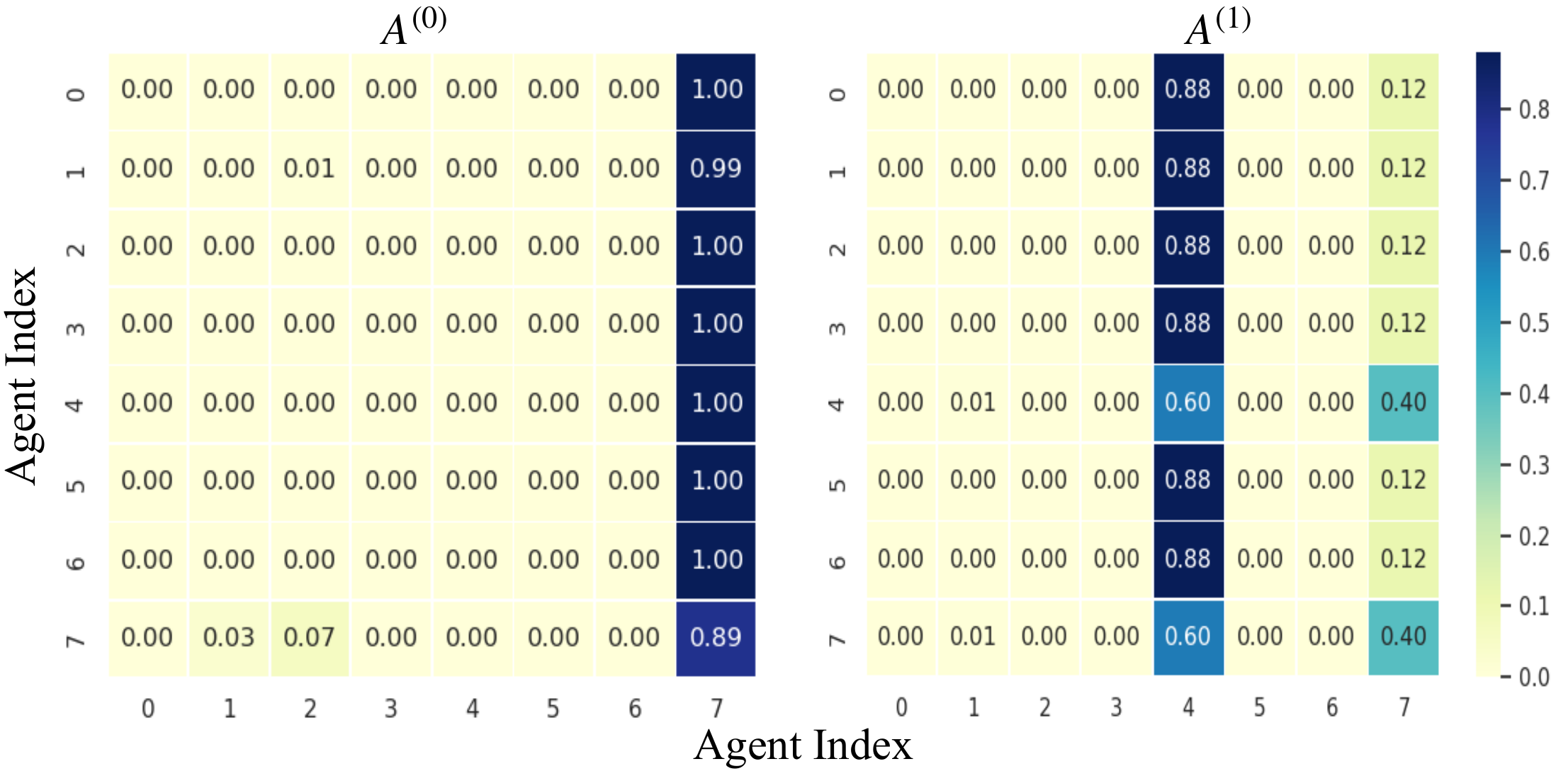}
		}
				\subfigure[The frontal slices with NTNNR.
		\label{fig-pp-TNNR-slice1}]
		{
			\includegraphics[width=0.5\linewidth]{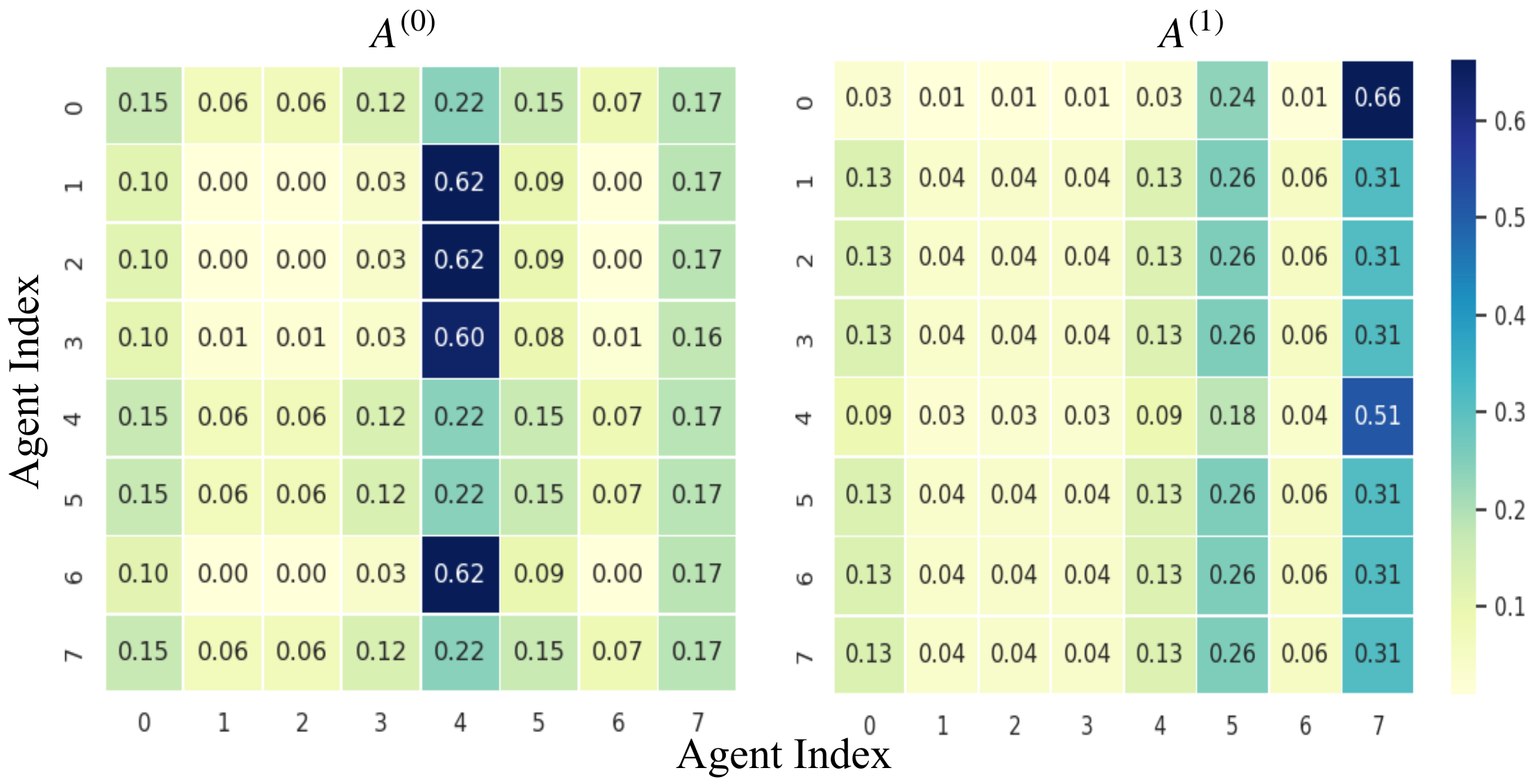}
		}
																										\caption{ Visualization of the adjacency tensors generated by two-attention GAT with TNNR and NTNNR.\label{fig:tensors}}
		
	\end{figure*}

	\section{Experimental Results}
	\noindent 
	In this part, we evaluate the performance of NTNNR in three widely-used scenarios: Predator-Prey, Traffic Junction, and StarCraft II Multi-Agent Challenge.
	
	In the mixed cooperative-competitive predator-prey scenario, we conduct ablation studies to show the effectiveness of NTNNR.
	We compare GAT with NTNNR with two baselines: vanilla GAT and applying our defined tensor nuclear norm without normalization (TNNR) to GAT. 
	In the cooperative traffic junction scenario, we compare our proposed message aggregation method, GAT with NTNNR, against a variety of widely used message aggregation methods, including averaging used in CommNet~\cite{sukhbaatar2016learning}, signature-based attention mechanism used in TarMAC~\cite{das2019tarmac}, GAT used in  GA-Comm~\cite{liu2020multi} and MAGIC~\cite{niu2021multi}, and GATv2~\cite{brody2022attentive}.
		Following the experimental setup in MAGIC, we utilize the two-layer GAT. The first layer contains two attention heads in the predator-prey scenario and four in the traffic junction scenario, while the second layer always contains one.
	For all methods, we uniformly adopt the REINFORCE~\cite{williams1992simple} with baseline as the training algorithm.

	StarCraft II Multi-Agent Challenge (SMAC)~\cite{whiteson2019starcraft} is a benchmark to evaluate various reinforcement learning works in recent years.
	Among them, we choose two state-of-the-art Comm-MARL methods, GA-Comm~\cite{liu2020multi} and DICG-CE-LSTM~\footnote{DICG-CE-LSTM is the communication-augmented version of DICG.}~\cite{li2021deep},
	and then integrate NTNNR with them. All results are obtained by averaging over three runs.

	\subsection{Predator-Prey}
	\noindent 
	Predator-Prey is one of the Multi-Agent Particle Environments~\cite{lowe2017multi}.
	In this scenario, we set 8 predators pursuing four fixed preys.
	To make the scenario mixed cooperative-competitive, two predators are required to be present in the grid cell of a prey for a successful capture.
	A predator obtains a reward of 0.3 if it captures a prey successfully.
	We set the maximum time steps to 30 and impose a step cost of 0.1.

			We set scaling hyper-parameters to $\beta_1 = 0.2, \beta_2 = 0.005$ for the two GAT layers respectively.
	Figure~\ref{fig-pp-reward} shows the average reward as the training epoch increases.
	Integrating NTNNR into GAT when aggregating messages could boost the training efficiency, and obtain the highest asymptotic performance.
	Two baselines need more time to explore emergent strategies, demonstrating that NTNNR incentivizes more efficient exploration and, finally, achieves better coordination.
	We also record the corresponding NTNN values in Figure~\ref{fig-pp-NTNNR-layer1} and \ref{fig-pp-NTNNR-layer2} respectively.
	We can observe that vanilla GAT keeps small NTNN values in both layers during the training stage, which suggests that all agents have homogeneous message aggregation strategies.
	In the early stage of training, GAT with NTNNR exhibits large NTNN values, encouraging agents to obtain more diverse message aggregation strategies and explore environments better.
	
	Compared to TNNR, NTNNR can maintain larger NTNN values in the second layer with the same scaling hyper-parameters. This is due to additional diversity among different attention heads.
	Note that even though GAT with NTNNR and with TNNR converges to similar NTNN values in the first layer, they have different message aggregation strategies.
	We visualize the adjacency tensors of two methods in similar states in Figure~\ref{fig:tensors}.
	Compared to TNNR, NTNNR can maintain diversity in inter and intra-frontal slices, indicating that normalization is critical for the regularizer when using the multi-head attention mechanism.

	\subsection{Traffic Junction}
	\noindent
	The second scenario we employ is cooperative.
	The hard-mode traffic junction scenario~\cite{sukhbaatar2016learning} consists of two-way intersecting routes on an $18 \times 18$ grids with four arrival points, and cars (agents) with one-grid limited vision, requiring communication to avoid collisions. 
	We set the maximum number of cars in the environment to 20 and
	the maximum time steps to 50.
	New cars get added to the environment with a probability of $0.05$.
		Success indicates that there are no collisions within an episode.
	The action space for each car is gas and break, and the reward consists of a step cost of $0.01$ and a collision penalty of $-10$.

	\begin{figure}[t]
												\centering
		\subfigure[ A test frame in Traffic Junction. 
		\label{fig-tf}]
		{
			\centering
			\includegraphics[width=0.48\linewidth]{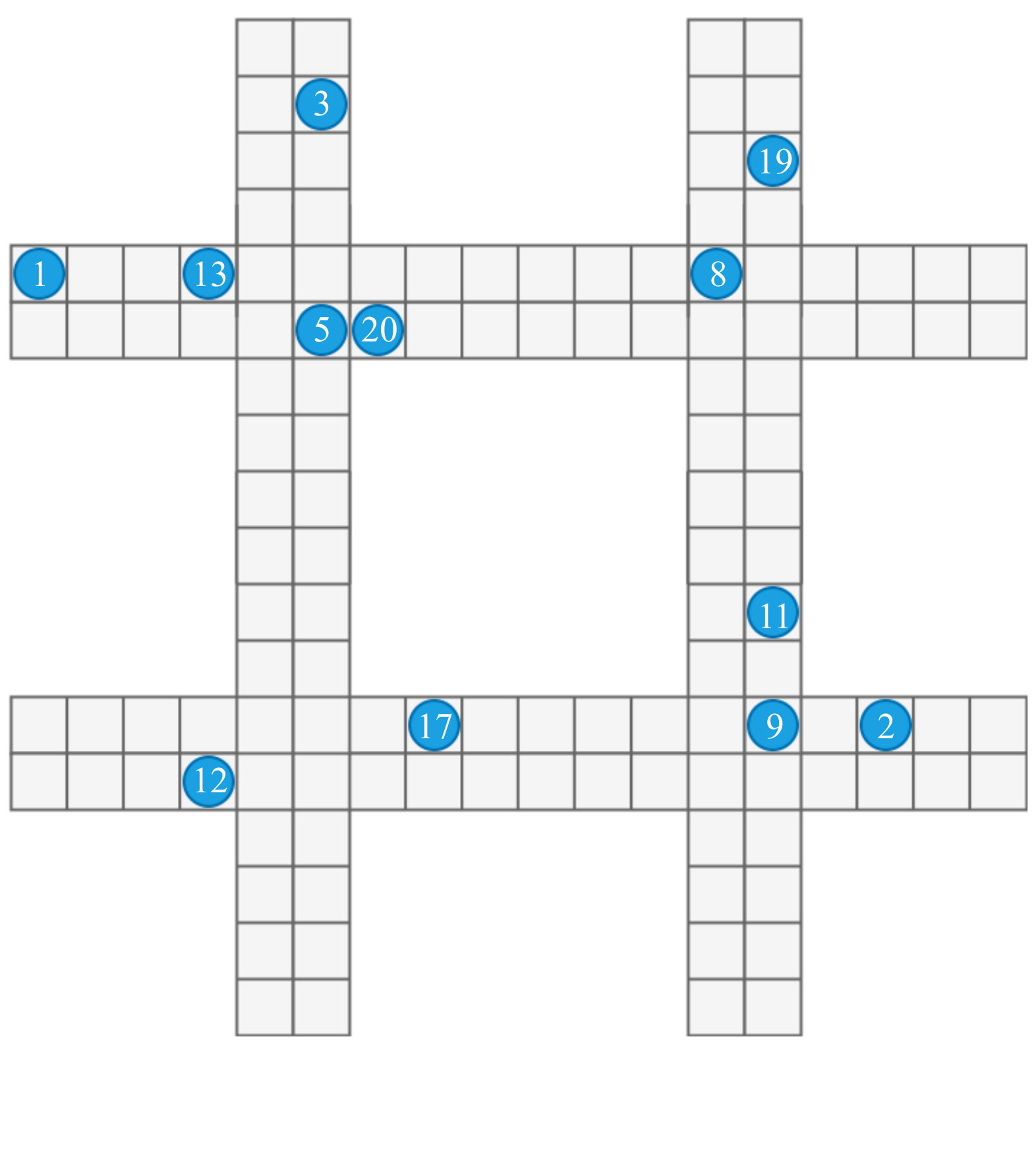}
		}
		\subfigure[Message aggregation strategies of agent $5$ and $9$
		\label{fig-tf-b}]
		{
			\includegraphics[width=0.45\linewidth]{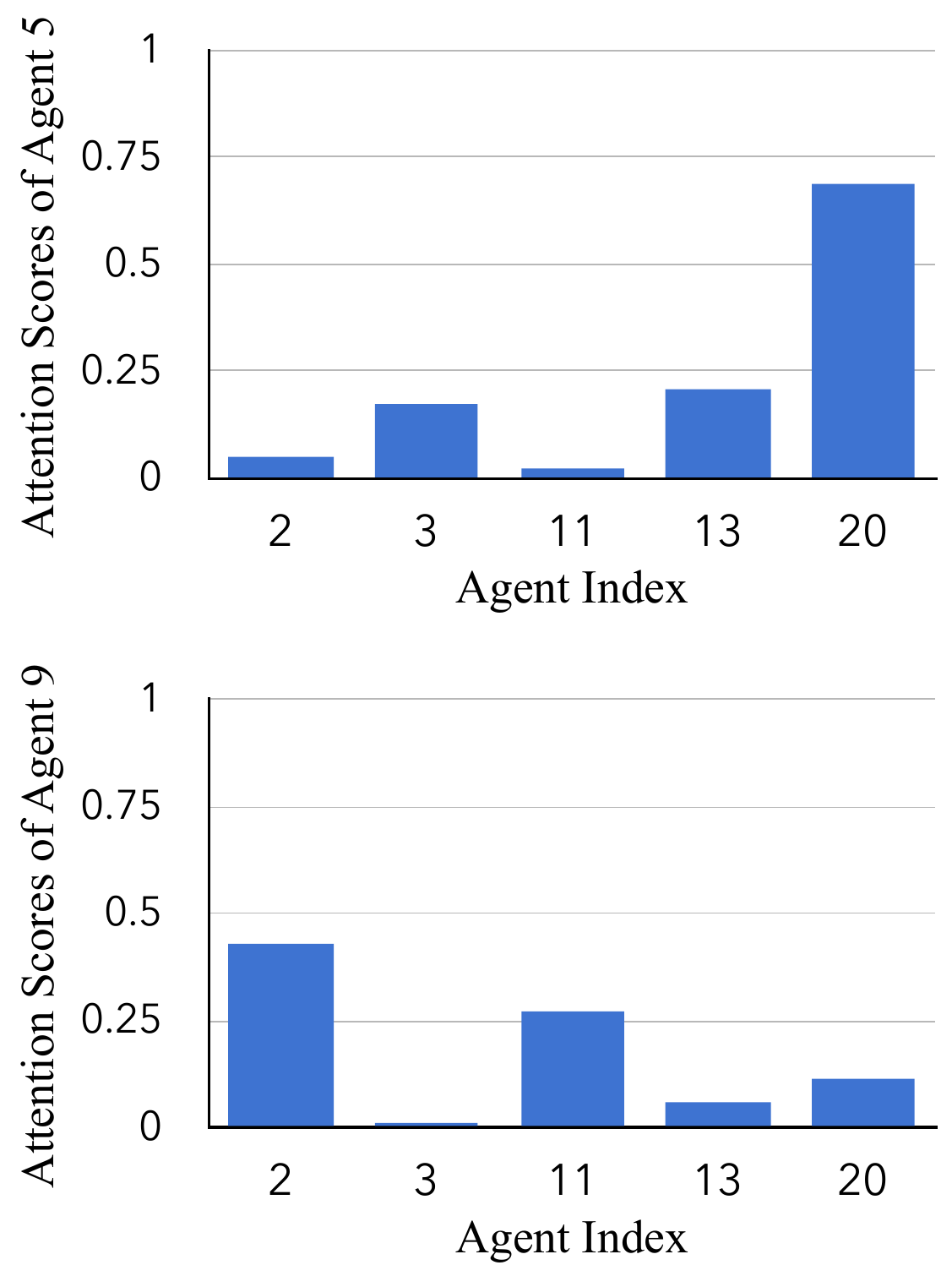}
		}
														
		\caption{The visualization of a time step in the Traffic Junction scenario (hard mode). \label{fig-tf-visual}}
		
	\end{figure}
	
	\noindent

	We set $\beta_1 = 0.01, \beta_2 = 0.005$ for the two GAT layers respectively.
		For the agents around different arrival points, NTNNR can encourage them to obtain diverse message aggregation strategies.
	The message aggregation strategies of agents are constantly changing at different time steps in an episode.
	In order to analyze the impact of NTNNR on strategies, we visualize the message aggregation strategies of two agents at one time step evaluated with the well-trained policy in Figure~\ref{fig-tf-visual}.
	It is observed that agent $5$ is at the upper left arrival point, while agent $9$ is at the downright arrival point.
	Intuitively, even though they can communicate, the messages are useless to each other.
	The distributions of representative attention scores for message aggregation are shown in Figure~\ref{fig-tf-b}.
	With NTNNR, agents $5$ and $9$ obtained diverse message aggregation strategies.
	This makes communication more efficient in the multi-agent system, avoiding unnecessary interference between unrelated agents.
	
\begin{figure}[ht]
		\centering
		\includegraphics[width=1\linewidth]{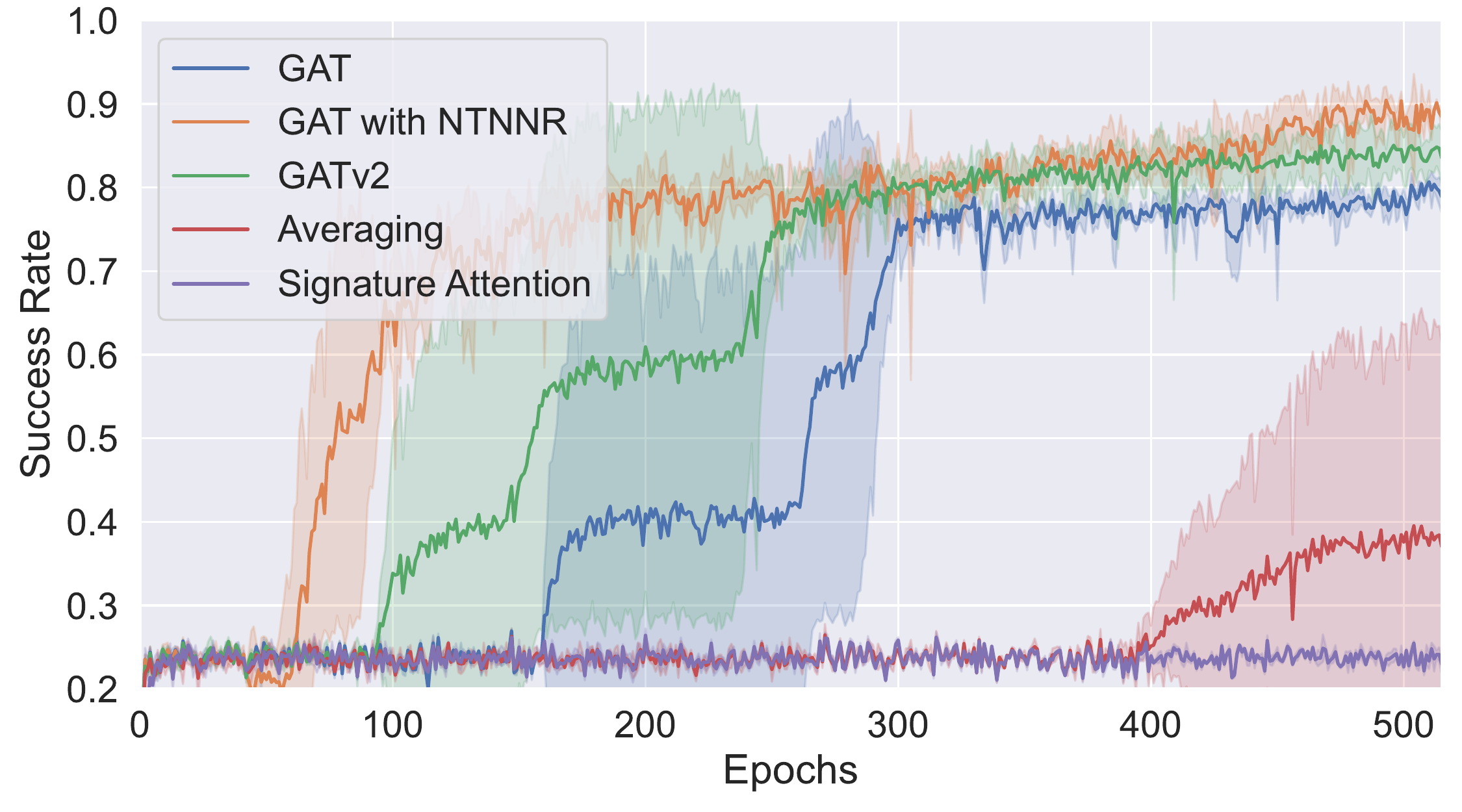}
		\caption{Success rates of various message aggregation approach in the traffic junction scenario.
			\label{fig-tf50-Success}}
		
	\end{figure}

	Figure~\ref{fig-tf50-Success} shows the success rate per epoch attained by various message aggregation methods.
	GAT with NTNNR is competitive when compared to other methods.
	Our method not only provides a higher success rate but also can be more sample efficient. We suppose the phenomenon attributes to efficient communication brought by NTNNR, where agents find optimal coordination faster.

	\subsubsection{The effect of the scaling hyper-parameters $\beta_l$:}
		\noindent 
	To further analyze the effect of the regularization weights of NTNNR, we evaluate the performance with different $\beta_1$ and $\beta_2$.
	We record the corresponding success rates in Table~\ref{table2}.
	From the first column and row, we can observe that utilizing NTNNR can significantly improve performance.
	
	\begin{table}[h]
		\centering
				\begin{tabular}{c|c|c|c|c|c}
			\hline 
			\diagbox{$\beta_1$}{$\beta_2$} &0& 0.001 & 0.005 & 0.01 & 0.02  \\  \hline
			0 &  0.77& 0.78 &0.86 & 0.83&0.84  \\ \hline
			0.005 & 0.87 & 0.84 & 0.82 &0.81 &0.89\\ \hline
			0.01 & 0.88 &  0.81& 0.91 &0.85 &0.83 \\  \hline
			0.02 &0.76  & 0.85 & 0.80 &0.85&0.87  \\  \hline
					\end{tabular}
		\caption{Success rates with different scaling hyper-parameters of NTNNR.}
		\label{table2}
	\end{table}
	
		We observe that the best performance achieved when $\beta_1=0.01, \beta_1=0.005$.
		Considering the first GAT layer contains four attention heads while the second layer only contains one, we recommend a larger regularization weight when the third dimension of the adjacency tensor is larger.
	Experiments in the predator-prey scenario also support this conclusion.
	Besides, it is observed that setting $\beta_1$ between $0.005$ and $0.01$ or setting $\beta_2$ between $0.001$ and $0.02$ can guarantee the performance improvement, showing acceptable robustness to the scaling hyper-parameters.

	\subsection{StarCraft II}

	\noindent 
	In this section, we evaluate our method on SMAC, a more complex benchmark.
	We want to show that NTNNR is general and easily integrated with existing graph-attention Comm-MARL methods, using the plug-and-play manner.
				We choose two state-of-the-art methods, GA-Comm and DICG-CE-LSTM, and apply NTNNR to them.
	The scaling hyper-parameter is set to $0.05$ and $0.005$, respectively.
			
	\begin{figure}[h]
														\centering
		\subfigure[2s3z
		\label{2s3z}]
		{
			\centering
			\includegraphics[width=0.47\linewidth]{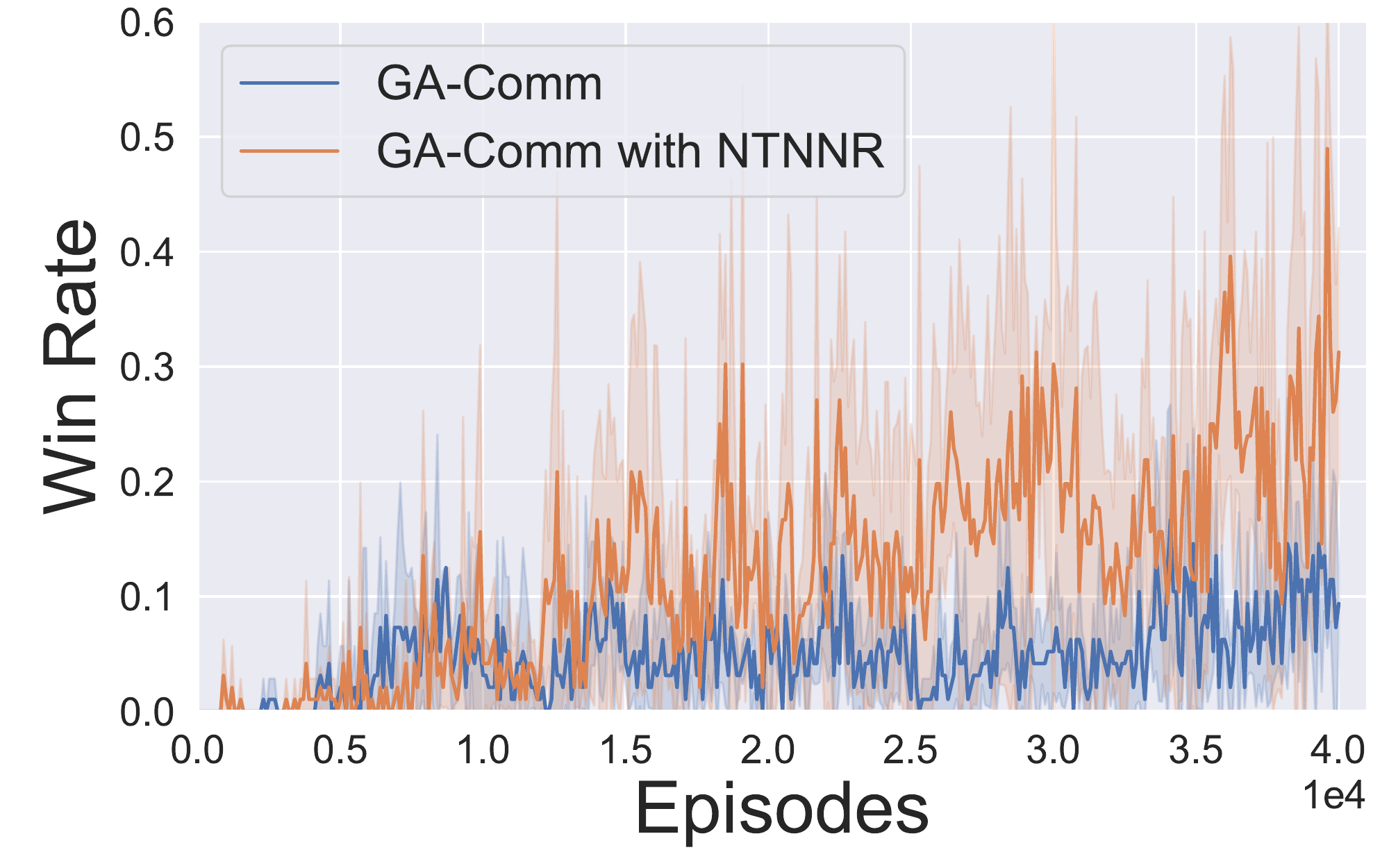}
		}
		\subfigure[3s5z
		\label{3s5z}]
		{
			\centering
			\includegraphics[width=0.47\linewidth]{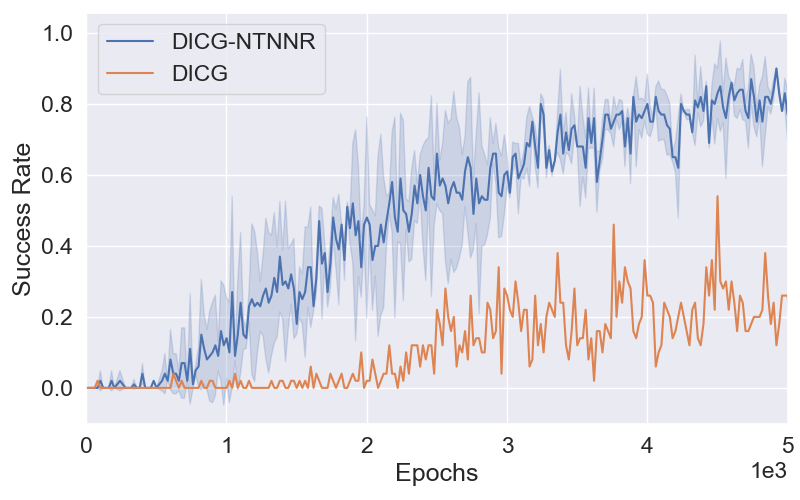}
		}
		\caption{Performance comparison of methods with NTNNR over their vanilla counterparts in SMAC maps. \label{SMAC}}
		
	\end{figure}

	\begin{figure}[ht]
		\centering
		\includegraphics[width=0.9\linewidth]{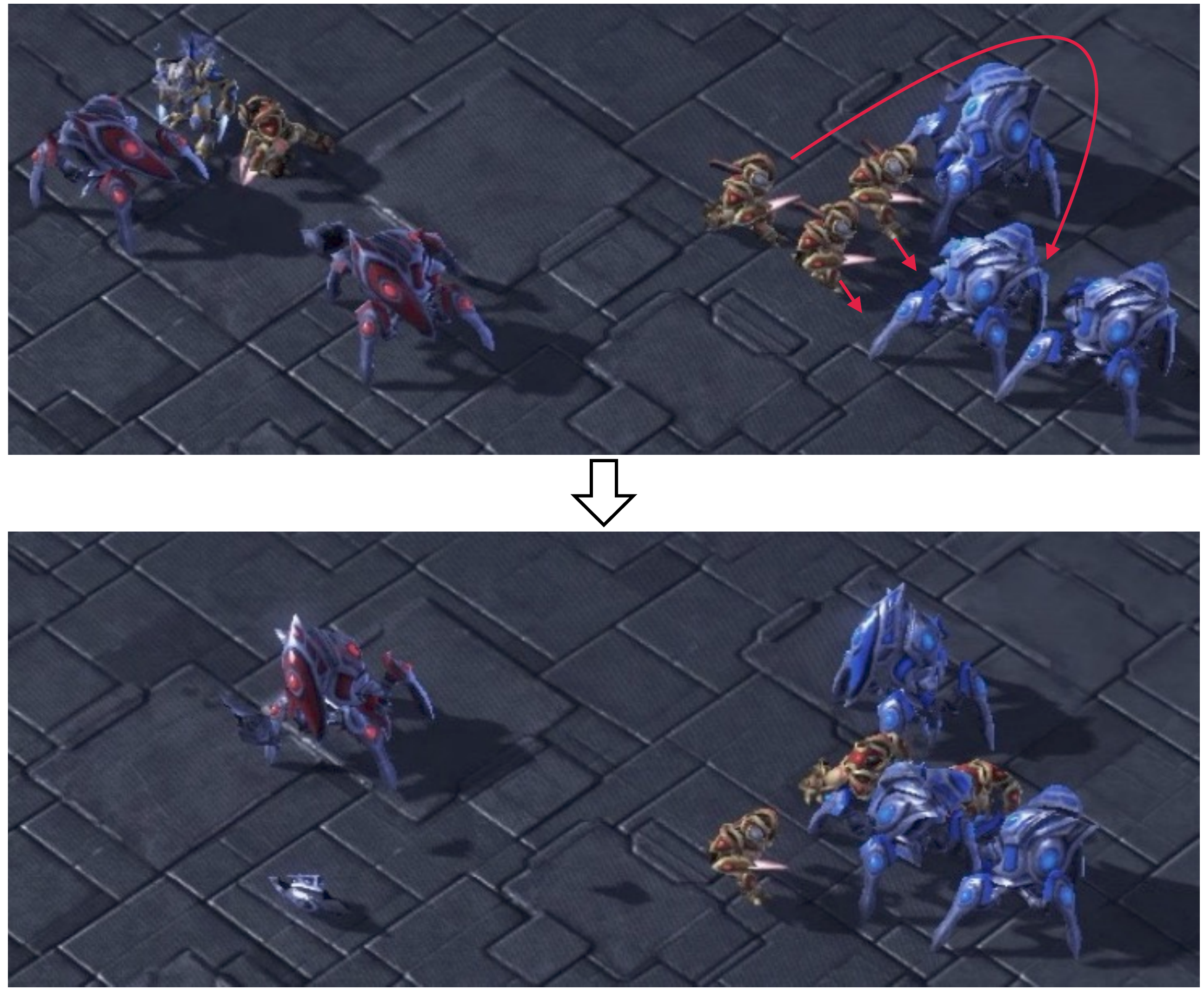}
		\caption{Visualization of the final strategies trained by DICG-CE-LSTM with NTNNR in the 3s5z map. We control the red stalkers and zealots.
			\label{visualization}}
		
	\end{figure}
	
	The average evaluation win rates are shown in Figure~\ref{SMAC}.
	Methods augmented by NTNNR achieve outstanding performance compared with their vanilla counterparts.
	We suppose the improvement is due to the emergent behaviors brought by the diverse message aggregation strategies.
		To better explain why our regularizer performs well, we further visualize the final trained strategies in Figure~\ref{visualization}.
	In this 3s5z map, three parameter-sharing zealots with similar observations can select diverse actions and finally surround the enemy stalkers to attack.
	The sophisticated coordination reflects the effectiveness of diverse message aggregation in Comm-MARL.

	\section{Conclusion}
	\noindent 
		In this paper, we present that the diversity of message aggregation in graph-attention Comm-MARL methods could be measured by the normalized tensor rank, and further define the corresponding nuclear norm to quantify the diversity.
	Then we propose a plug-and-play regularizer named NTNNR, to actively enrich the diversity of message aggregation. 
	Experiments show that GAT with NTNNR can provide superior performance and better training efficiency compared to existing message aggregation methods.
	Furthermore, NTNNR can be easily applied to existing graph-attention Comm-MARL methods and improve their performance.
	
	Assuredly, our method has some limitations. In some multi-agent coordination tasks with core agents, overly diverse message aggregation may be unreasonable.
	Therefore, NTNNR may not achieve significant performance improvements in these cases.
	In future work, we plan to quantify the diversity upper bound for multi-agent systems.

	\bibliography{aaai23}
	
\clearpage
\appendix
\section{Analysis of Diversity Maintained by NTNNR}
\noindent
In this paper, we define the Normalized Tensor Rank and the Normalized Tensor Nuclear Norm (NTNN). 
Enlarging NTNN can enrich the diversity of tensor $\mathcal{A}$ from both the frontal slice view and the mode-3 fiber view.
For better comprehension of the effect, we will take a toy example.
Suppose there are two agents in the multi-agent system, and the multi-head attention mechanism contains two heads, i.e., $N=2, K=2$.
In this case,  $\mathcal{A} \in \mathbb{R}_+^{2 \times 2 \times 2}$ could be expressed as:
\begin{equation}
	\mathbf {A }^{(0)} = \begin{bmatrix}  x_0 & 1-x_0 \\   y_0 & 1-y_0\end{bmatrix}, \ \ 
	\mathbf {A }^{(1)} = \begin{bmatrix}  x_1 & 1-x_1 \\   y_1 & 1-y_1\end{bmatrix}.
\end{equation}
where $x_0$, $y_0$, $x_1$ and $y_1$ are variables. 

Applying normalization to $\mathcal{A}$ along the 3-rd way and transforming the tensor to the block diagonal matrix form, we have:
\begin{equation}
	\mathbf {\hat A} = \begin{bmatrix}  
		\hat x_0 & 1-\hat x_0 \\   \hat y_0 & 1-\hat y_0  \\
		& &  \hat x_1 & 1-\hat x_1 \\   & &  \hat y_1 & 1-\hat y_1 \\
	\end{bmatrix}.
\end{equation}

To obtain the singular values, we calculate the eigenvalues of $\mathbf {\hat A}  \mathbf {\hat A}^T$ as follows:
\begin{equation}
	| \mathbf {\hat A}  \mathbf {\hat A}^T - \lambda  \mathbf {\hat I} | = 0.
	\label{eqa}
\end{equation}

We denote the four singular values as $\sigma_0, \sigma_1, \sigma_2$, and $\sigma_3$ respectively.
With the properties of block diagonal matrix, we solve Equation~\ref{eqa} and have:

\begin{equation}
	\left\{
	\begin{array}{ll}
		\sigma_0^2 + \sigma_1^2 = 2(\hat x_0^2 - \hat x_0 + \hat y_0^2 - \hat y_0 +1), \\
		\sigma_0^2 \times \sigma_1^2  = (\hat y_0- \hat x_0) ^2, \\
		\sigma_2^2 + \sigma_3^2 = 2(\hat x_1^2 - \hat x_1 + \hat y_1^2 - \hat y_1+1), \\
		\sigma_2^2 \times \sigma_3^2  = (\hat y_1- \hat x_1) ^2, \\
		\hat x_0+ \hat x_1 = 1, \\
		\hat y_0+ \hat y_1 = 1.
	\end{array}
	\right.
\end{equation}

The normalized tensor nuclear norm is the sum of the singular values of $\mathbf {\hat A}$, which is calculated as follows:
\begin{equation}
	\begin{aligned}
		\Vert \mathcal{A} \Vert _ *&	=\frac{1}{K}\Vert \mathbf {\hat A} \Vert _ * =\frac{1}{K} ( \sigma_0 + \sigma_1 + \sigma_2 + \sigma_3)\\ 
		&= \frac{1}{K} (\sqrt{(\sigma_0 + \sigma_1)^2} + \sqrt{(\sigma_2 + \sigma_3)^2}) \\ 
		&= \sqrt{\hat x_0 + (1-\hat x_0)^2 + \hat y_0 + (1-\hat y_0)^2 + 2|\hat y_0-\hat x_0|}
		\\ s.t. \ \ \  & \hat x_0 + \hat x_1 = 1, \ \  \hat y_0 + \hat y_1 = 1.
	\end{aligned}
	\label{eq:toy1}
\end{equation}


From Equation~\ref{eq:toy1}, $\Vert \mathcal{A} \Vert _ *$ would reach the maximum solution when:
\begin{equation}
	\begin{aligned}
		& \mathbf {A }^{(0)} = \begin{bmatrix}  0 & 1\\   1 & 0\end{bmatrix} \ \ 
		\mathbf {A }^{(1)} = \begin{bmatrix}  1 & 0   \\  0 & 1\end{bmatrix} , \ \ \text{or}
		\\ & \mathbf {A }^{(0)} = \begin{bmatrix}  1 & 0  \\  0 & 1\end{bmatrix} \ \ 
		\mathbf {A }^{(1)} =  \begin{bmatrix}  0 & 1 \\   1 & 0\end{bmatrix}.
	\end{aligned}
	\label{max-example}
\end{equation}

Therefore, we demonstrate that NTNNR tries to maintain the diversity of the adjacency tensor in both the frontal slice view and the mode-3 fiber view.

\section*{Code}
\noindent
Our code will be released publicly to enhance the reproducibility.
We use the following open-source repositories for baselines:

\begin{itemize}
	\item MAGIC code: \href{https://github.com/CORE-Robotics-Lab/MAGIC}{https://github.com/CORE-Robotics-Lab/MAGIC}
	\item GA-Comm code: \href{https://github.com/starry-sky6688/MARL-Algorithms}{https://github.com/starry-sky6688/MARL-Algorithms}
	\item DICG-CE-LSTM code: \href{https://github.com/sisl/DICG}{https://github.com/sisl/DICG}
\end{itemize}



\section{Implementation Details}
\label{Sec:PG}
\noindent
Our implementation is on a desktop machine with one Intel i9-12900K CPU and one NVIDIA RTX3080 GPU. All the methods in the same scenario are run for the same number of total environment steps (or episodes) and the same number of iterations.

\par
\begin{table}[h]
	\centering
	\caption{\footnotesize{Hyper-parameters in the the Predator-Prey and Traffic Junction Scenarios.}}
	\label{hyper}
	\begin{tabular}{l|c|c}
		\hline Parameter & \makecell[c]{Predator\\-Prey} & \makecell[c]{Traffic\\Junction} \\
		\hline 
		Number of processes & 16&16 \\
		Epoch size & 10&10 \\
		Hidden units for LSTM encoder& 128 &128\\
		Learning rate & 0.001 &0.001\\
		\makecell[l]{Number of attention heads\\(the first GAT layer)} & 2 & 4 \\
		\makecell[l]{Number of attention heads\\(the second GAT layer)} & 1 & 1 \\
		Hidden units of each attention head  & 32& 32 \\
		Scaling hyper-parameter $\beta_1$   & 0.2 & 0.01 \\
		Scaling hyper-parameter $\beta_2$   & 0.005 & 0.005 \\
		\hline
	\end{tabular}
\end{table}
\par


In the the Predator-Prey and Traffic Junction Scenarios, we distribute the training over 16 threads and each thread runs batch learning with a batch size of 500. The threads share the parameters $\boldsymbol{\theta}$ of the policy network and update synchronously. We use RMSProp as the optimizer.
Table \ref{hyper} shows the details of our neural network and other hyper-parameters.

In the StarCraft II scenario, we follow the original implementations of the selected Comm-MARL methods.
We adopt their network structures and hyper-parameter settings, except changing the number of attention heads to $4$ and the hidden units of each attention head to $32$ for GA-Comm.


\section{Training Algorithms}
Most of existing Comm-MARL methods utilize policy gradient methods with parameter sharing.
Then the joint policy can be factorized as:

\begin{equation}
	\pi_{\boldsymbol{\theta}}(\mathbf{u}^t | \mathbf{o}^t, \mathbf{m}^t) = \prod \limits_{i=0}^N \pi_{\boldsymbol{\theta}}(u_i^t | o_i^t, m_{j \neq i}^t)
\end{equation}

Augmented with communication, the environment and other agents' policies can be seen as stable for a agent.
Agents simultaneously select actions according to local observations and communication messages.
So various single-agent policy-gradient methods can be utilized as the training algorithms for Comm-MARL.

All methods used in the predator-prey and traffic junction scenarios adopt the REINFORCE~\cite{williams1992simple} with baseline as training algorithms.
In this case, Equation 9 of the manuscript can be written as:

\begin{equation}
	\begin{aligned}
		\nabla_{\boldsymbol{\theta}}	{L}_{RL}(\boldsymbol{\theta})  =  \mathbb{E}_{i,t} [\nabla_{\boldsymbol{\theta}}log \  \pi_{\boldsymbol{\theta}}& (u_i^t | o_i^t, m_{j \neq i}^t)  \psi^t_i ],
	\end{aligned}
\end{equation}
where $\psi^t_i=r^t_i -V(o_i^t, m_{j \neq i}^t)$ is the advantage function. $V(o_i^t, m_{j \neq i}^t)$ is the value function.

GA-Comm adopts the REINFORCE algorithm. The Equation 9 of the manuscript can be written as:
\begin{equation}
	\begin{aligned}
		\nabla_{\boldsymbol{\theta}}	{L}_{RL}(\boldsymbol{\theta})  =  \mathbb{E}_{i,t} [\nabla_{\boldsymbol{\theta}}log \  \pi_{\boldsymbol{\theta}}& (u_i^t | o_i^t, m_{j \neq i}^t)  \sum ^{T}_{t ^ \prime =t} \gamma ^{t ^ \prime -t}r^{t ^ \prime}_i ],
	\end{aligned}
\end{equation}
where $T$ is the maximum time steps for agents to interact with the environment.

DICG-CE-LSTM adopts the clipped PPO~\cite{schulman2017proximal} algorithm. Then the Equation 9 of the manuscript has the following form:
\begin{equation}
	\begin{aligned}
		\nabla_{\boldsymbol{\theta}} 	{L}_{RL}&(\boldsymbol{\theta})  =  \mathbb{E}_{i,t} [\nabla_{\boldsymbol{\theta}}
		min(\frac{\pi_{\boldsymbol{\theta}}(u_i^t | o_i^t, m_{j \neq i}^t)}{\pi_{old}(u_i^t | o_i^t, m_{j \neq i}^t)}\psi^t_i,\\
		&clip(\frac{\pi_{\boldsymbol{\theta}}(u_i^t | o_i^t, m_{j \neq i}^t)}{\pi_{old}(u_i^t | o_i^t, m_{j \neq i}^t)},1-\epsilon,1+\epsilon)
		)\psi^t_i)],
	\end{aligned}
\end{equation}
where $\epsilon$ is the hyper-parameter for clip, and $\pi_{old}$ is the policy of the last update iteration.
																																																																																																																																																																																																																															
	\end{document}